\documentclass[preprint,showpacs]{revtex4}
\usepackage{epsfig}
\usepackage{graphicx}
\usepackage{bm}
\newcommand{\bmath}[1]{\mbox{\boldmath{${#1}$}}}

\newcommand{\rd}{\textrm{d}}
\begin{document}
\title{A remark on the Primakoff effect }
\author{G\"oran F\"aldt}\email{goran.faldt@fysast.uu.se} 
\affiliation{ Department of physics and astronomy, 
Uppsala University,
 Box 516, S-751 20 Uppsala,Sweden }

\begin{abstract}
 The coherent-nuclear reaction $ a +{\rm A}\rightarrow a^{\star} +{\rm A}$ 
 is in the small-angle region dominated by the one-photon-exchange 
 mechanism, 
often referred to as the Primakoff effect. In this region  
information about 
the electromagnetic  decay  $ a^{\star} \rightarrow a +\gamma $ can
be obtained.
Well-known examples are the two-photon decays of the pi- and eta-mesons. 
Also decays of charged hadrons  can be studied.
For charged hadrons the one-photon-exchange  amplitude comes with a 
 Coulomb-phase factor and a Coulomb-form factor, which depend on the
ratio between transverse- and  logitudinal-momentum transfers,
the latter being fixed.  At the peak of the cross-section distribution,
where the two momentum transfers are equal, 
the form factor could cut down the cross-section value by as much as 40\%.
 Consequently,
 a determination of a radiative-decay rate that relies on the peak value 
 becomes sensitive to a proper treatment of the Coulomb-form factor.
 \end{abstract}
\maketitle
%
%
%
Many radiative-hadronic-decay rates have been determined through  the Primakoff effect.
The theory for the Primakoff effect, i.e.~coherent-nuclear 
production in the one-photon-exchange approximation,
 is laid out in ref.\cite{Prikoff}. 
 In the case of charged hadrons the formalism was extended in  ref.\cite{GFbas} 
 to include elastic-Coulomb scattering.
 There, also the appropriate treatment
 of the coherent-nuclear background was described. The point we would
like to make here is that starting from the formalism 
of \cite{GFbas}, 
 the amplitude for a point-like nucleus can be calculated analytically.
 We shall then see that for charged hadrons the point-like-Coulomb
  amplitude is adorned with a form factor, extremely important
in the vicinity the Primakoff peak.

The amplitude of the proton reaction, $ a(k) +p(p)\rightarrow a^{\star}(k') +p(p')$,
can in the one-photon-exchange approximation be written as
\begin{equation}
F_p(\mathbf{q}_{\bot},q_{\|})= 2\alpha 
\frac{\mathbf{c}\cdot\mathbf{q}_{\bot}}{\mathbf{q}_{\bot}^2+q_{\|}^2} ,
\label{Prim-f-prot}
\end{equation}
where $\mathbf{c}$ is a vector in the impact-parameter plane, defined as the plane orthogonal
to the incoming momentum $\mathbf{k}$. The components of the 
momentum transfer $\mathbf{q}$ are defined as
 $(\mathbf{q}_{\bot},q_{\|})= \mathbf{k}'-\mathbf{k}$. 
 The longitudinal component $q_{\|}$ 
 is fixed, and given by the expression
\begin{equation}
q_{\|}=-(m_{a^{\star}}^2-m_a^2)/2k.
\end{equation}

The Coulomb-production potential is proportional to $\mathbf{c}\cdot\mathbf{e}(\mathbf{r})$, 
where $\mathbf{e}(\mathbf{r})$ represents the electric field of the proton. The 
proton amplitude of Eq.(\ref{Prim-f-prot}) is in fact the Born approximation of this potential,
i.e.~
\begin{equation}
F_p(\mathbf{q})= \frac{-\alpha}{2\pi i}\int\rd^3r e^{-i\mathbf{q}\cdot\mathbf{r}}\ 
\frac{\mathbf{c}\cdot\mathbf{r}}{r^3}   , \label{Protampl}
\end{equation}
with $\mathbf{r}/{r^3}$ the electric field of the proton-point-charge distribution.

For a nuclear target the  production takes place in the Coulomb field of the
nucleus. If we simplify to a point-like nucleus this implies a multiplication
of the proton amplitude
by a factor of $Z$. However, if the projectile is charged  we must also take into 
account the distortion of the trajectory due to the elastic-Coulomb scattering.
This is done by introducing a Coulomb-phase factor \cite{RJG},
\begin{equation}
	e^{i\chi_C(b)}=\left( \frac{2a}{b}\right)^{i\eta}
\end{equation}
where $a$ is the cutoff radius in the Coulomb potential.
For a negatively charged projectile
\begin{equation}
	\eta =2 Z\alpha/v  .\label{etadef}
\end{equation}
The velocity $v$ can be put to unity. Thus, 
Eq.(\ref{Protampl}) generalized to nuclear scattering becomes
\begin{equation}
F_Z (\mathbf{q})= \frac{-Z\alpha}{2\pi i}\int \rd^3r e^{-i\mathbf{q}\cdot\mathbf{r}}
	 \      \frac{\bmath{c}\cdot\bmath{r}}{r^3}
	       \left( \frac{2a}{b}\right)^{i\eta} . \label{study-int-2}	
\end{equation}
Integration over the $z$-variable yields a modified Bessel function,
but also the integration over the impact parameter can be 
performed analytically \cite{FT}.
We write the result as
\begin{equation}
F_Z (\mathbf{q}) =  \mathbf{c}\cdot\mathbf{q}\ F_{C}(\mathbf{q}) ,
 \label{Coul-amp-fact}	
\end{equation}
splitting off the off-shell elastic-Coulomb-scattering amplitude 
$F_{C}(\bmath{q})$, 
\begin{equation}
	F_{C}(\bmath{q})= 
	\frac{2Z\alpha (aq)^{i\eta}e^{i\sigma_\eta }}{\bmath{q}^2}\ 
	 h_C(\bmath{q}) \label{FF-with-phase}
\end{equation}
with $\eta$ defined in Eq.(\ref{etadef}) and
\begin{equation}
	\sigma_\eta =2 \arg \Gamma(1-i\eta/2) .
\end{equation}
The extracted phase factors in Eq.(\ref{FF-with-phase}) 
are the same as in  elastic-Coulomb
scattering, except that  now 
\begin{equation}
	q=\sqrt{\bmath{q}_{\bot}^2 +  q_{\|}^2} .
\end{equation}
In high-energy-elastic scattering the longitudinal-momentum transfer 
$q_{\|}$ vanishes. In that case $q$  of Eq.(\ref{FF-with-phase})
is simply ${q}_{\bot}$.
The Coulomb  form factor $h_C(\bmath{q})$ emerges as \cite{FT}
\begin{equation}
	 h_C(\bmath{q})  =
    \Gamma(2-i\eta/2) \Gamma(1+i\eta/2)
	    F(i\eta/2, 1-i\eta/2 ;2; \frac{q_{\bot}^2}{q_{\bot}^2+q_{\|}^2}) .
	      \label{def-FF-coul}
\end{equation}

In elastic scattering the longitudinal-momentum transfer vanishes,
and $h_C(q_{\bot},q_{\|}=0)=1$ as expected. In Coulomb production with
neutral projectiles
there is no Coulomb scattering and consequently no Coulomb-form 
factor, and for charged projectiles $h_C(\mathbf{q})=1$ 
for $q_{\bot}\gg q_{\|}$, i.e.~for transverse-momentum transfers 
far beyond the Primakoff peak.

The form factor  $h_C(\bmath{q})$ is important only in the peak region and there, it is 
unimportant for light nuclei but extremely important for heavy nuclei.
This is illustrated in Fig.~1a where we have plotted the 
 spin-averaged cross section with and without  form factor for the lead
 nucleus. The solid line traces the  cross-section distribution ignoring the
form factor, and the dashed line the distribution with the form factor 
$h_C(\bmath{q})$. 
The  normalization is chosen so that the distribution without 
form factor peaks at one.  The variable $\zeta$ along
 the $x$-axis measures the ratio $\zeta={q}_{\bot}/q_{\|}$.

%
\begin{figure}[h]\label{Ratio-fig}\begin{center}
\scalebox{0.40}{\includegraphics{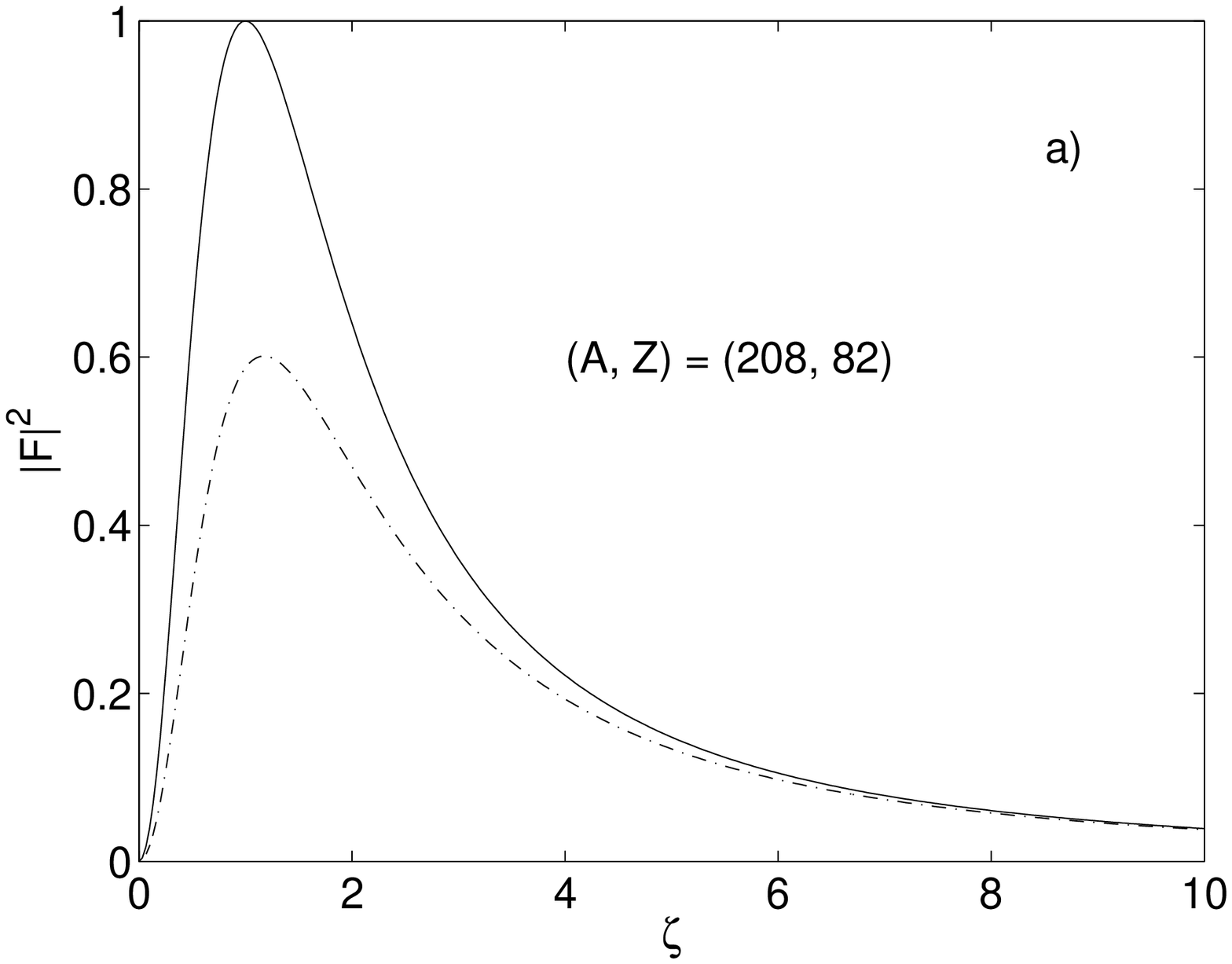} 
\quad\qquad\includegraphics{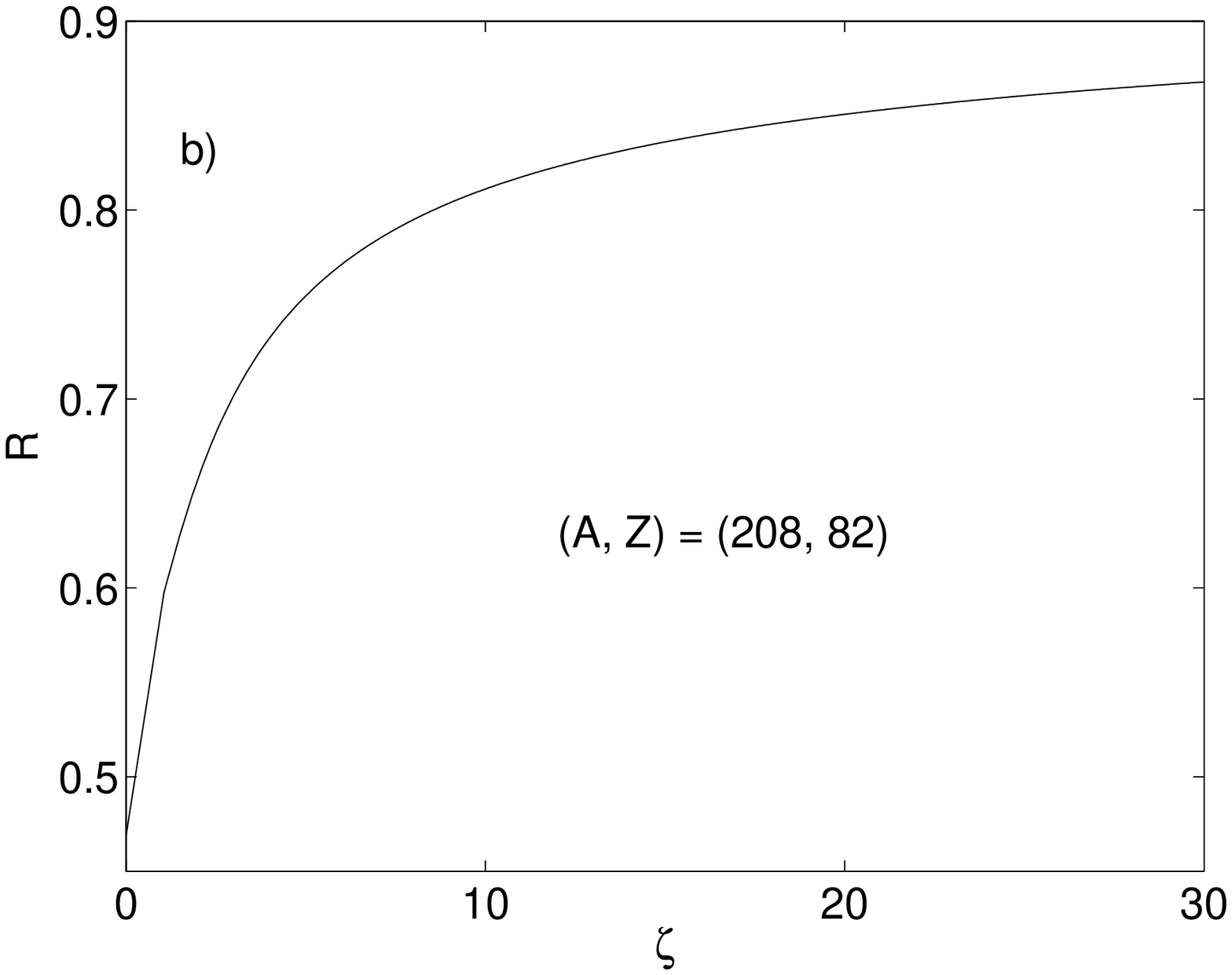}}
\end{center}
\caption{a) Cross-section distribution in the peak region. Solid line
calculated without
and dash-dotted line with Coulomb-form factor. The variable $\zeta=q_{\bot}/q_\|$.
b) Ratio $R$ of cross-section distributions integrated up to 
transverse-momentum transfer $\zeta$. In the numerator of $R$, the distribution with
and in the denominator the distribution without Coulomb-form factor.}
\end{figure}
%

Obviously, for heavy nuclei, the form factor is very important in the peak region.
 If it is possible experimentally to map out the peak 
that is advantageous. But in many situations the whole peak and even 
more is contained in the first momentum-transfer bin.
In such a case it is more useful to look at the ratio of 
the cross-section distributions
integrated out to some value of momentum transfer. In  Fig.~1b the 
integrated-cross-section ratio 
\begin{equation}
	R(\zeta)= \left[ \int_0^{\zeta^2} \rd y  \frac{y}{(1+y)^2}\big|h_C(y)\big|^2\right]
	 \bigg/ \left[
	\int_0^{\zeta^2} \rd y  \frac{y}{(1+y)^2}\right]
	\end{equation}
is graphed. As before $\zeta={q}_{\bot}/q_{\|}$. 
 %
We conclude that for a lead nucleus the Coulomb-form factor also dominates the integrated cross-section value far beyond the the peak. 

In this note we have considered the contribution from the
point-like-Coulomb-nuclear-charge distribution. The residual term due to the
extension of the charge is  easy to calculate numerically
once we have extracted analytically the point-charge amplitude. This additional term
is small in the region of the peak. Another neglected contribution 
is the coherent-nuclear-production contribution. It is 
negligible at high energies but can be important at low energies. 
These two terms are preferrably handled with the methods of ref.\cite{FT}.

The Coulomb-form factor is always important in the peak region. Nevertheless,
there are experiments at high energies where it  seems to have been 
ignored \cite{Hust,Ant}, but it is not clear how that omission affects the radiative-decay
rates extracted. On the other hand, there are also experiments both at high
\cite{Jens} and low energies \cite{Bemp} where coherent-nuclear production and
Coulomb-scattering effects, including the extension of the nuclear-charge distribution,
are fully treated in the formalism of \cite{GFbas}. Those analyses would have
been much simplified with the analytic form factor of Eq.(\ref{def-FF-coul})
as a starting point.


\end{document}